\documentstyle[nato,epsf]{crckapb} 

\begin{opening}
\title{FINE-GRID SIMULATIONS OF THERMALLY ACTIVATED \protect\\
       SWITCHING IN NANOSCALE MAGNETS}

\author{G. Brown$^{1,2}$}
\author{M.A.\ Novotny$^1$}
\author{P.A.\ Rikvold$^{1,2,3}$}
\institute{$^1$School of Computational Science and Information Technology, \protect\\
           $^2$Center for Materials Research and Technology, and\protect\\
           $^3$Department of Physics\protect\\
           Florida State University,
           Tallahassee, FL 32306-4120, USA}

\end{opening}

\runningtitle{FINE-GRID SIMULATIONS}

\begin{document}


\begin{abstract}
Numerical integration of the Landau-Lifshitz-Gilbert equation with
thermal fluctuations is used to study the dynamic response of
single-domain nanomagnets to rapid changes in the applied magnetic
field. The simulation can resolve magnetization patterns within
nanomagnets and uses the Fast Multipole method to calculate
dipole-dipole interactions efficiently. The thermal fluctuations play
an essential part in the reversal process whenever the applied field
is less than the zero-temperature coercive field. In this situation
pillar-shaped nanomagnets are found to reverse through a local curling
mode that involves the formation and propagation of a domain
wall. Tapering the ends of the pillars to reduce pole-avoidance effects
changes the energies involved but not the fundamental process. The
statistical distribution of switching times is well described by the
independent nucleation and subsequent growth of regions of reversed
magnetization at both ends of the pillar.
\end{abstract}

Magnetic nanoparticles are important components of
nanotechnology. Previously, Ising models, which describe only highly
anisotropic materials, have been used to understand thermally
activated switching in single-domain magnets \cite{RIKV}.
Single-domain magnets with dimensions of only a few nanometers are
being manufactured and measured, for instance, using scanning
microscopy techniques \cite{WIRTH}.  We have chosen these nanomagnets
as a specific system in which to investigate magnetization switching
with more realistic computational models of magnetic materials.

Here magnetic materials are modeled by position-dependent
magnetization density vectors ${\bf M}({\bf r})$ with fixed length
$M_s$, that precess around the local field ${\bf H}({\bf r})$
according to the Landau-Lifshitz-Gilbert equation
\cite{BROWN63,AHARONI}
\begin{equation}
\frac{ {d} {\bf{M}}({\bf{r}}) }
     { {d} {t} }
 =
   \frac{ \gamma_0 }
        { 1+\alpha^2 }
   {\bf{M}}({\bf{r}})
 \times
 \left[
   {\bf{H}}({\bf{r}})
  -\frac{\alpha}{M_s} {\bf{M}}({\bf{r}}) \times 
                      {\bf{H}}({\bf{r}})
 \right]
\;,
\end{equation}
where $\gamma_0 = 1.76\times10^7\,{\rm Hz/Oe}$ is the electron
gyromagnetic ratio and $\alpha$ is a phenomenological damping
parameter. The value $\alpha=0.1$ was chosen to give an underdamped
system. The saturation magnetization $M_s=1700\,{\rm emu/cm^3}$ and
the exchange length $\ell_x=2.6\,{\rm nm}$ were chosen to match those
of bulk iron. Numerical integration of Eq.~(1) was carried out using a
finite-differencing scheme with $\Delta r = 1.5\,{\rm nm}$ and $\Delta
t = 5\times10^{-5}\,{\rm ns}$. The local field includes exchange
interactions, dipole-dipole interactions, and thermal
fluctuations. Details of the numerical approach will appear in
Ref.~\cite{MMAG}.

Here we consider nanomagnets with square cross sections of side
$9\,{\rm nm}$. One geometry has flat ends, and is $150\,{\rm nm }$
along the $z$-direction \cite{JAP2000}. A simulated hysteresis loop at
$T=0\,{\rm K}$ with a period of $4\,{\rm ns}$ for this system is shown
in Fig.~1. There is no crystalline anisotropy included in the model,
but the shape anisotropy is quite strong, giving a coercive field of
about $1875\,{\rm Oe}$ \cite{MMAG}. Even for these extremely fast
loops, for a significant amount of time the magnet experiences a field
just below the coercive field, and thermal fluctuations can carry the
magnetization through the configuration associated with the energy
barrier. Some hysteresis loops simulated at $T=100\,{\rm K}$ are also
shown in Fig.~1 to illustrate the extent of this effect.

\begin{figure}
\vspace{6.0truecm}
\includegraphics{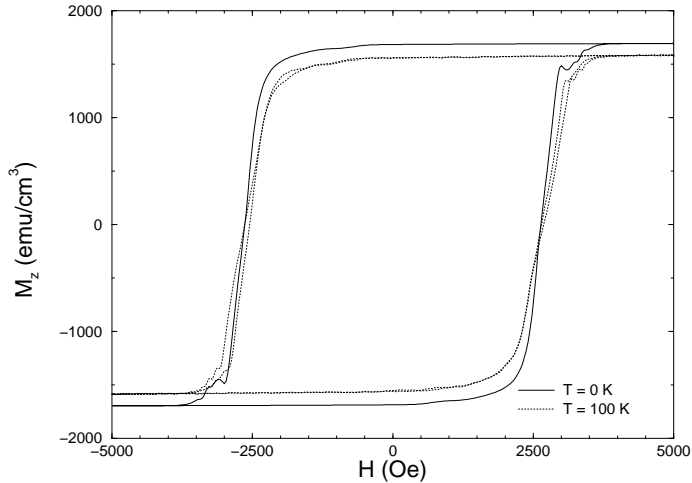}
\caption[] {Hysteresis loops of period $4\,{\rm ns}$ for flat-ended
nanoscale pillars described in the text. One is at $T=0\,{\rm K}$
(solid) and two at $T=100\,{\rm K}$ (dotted).}
\end{figure}

\begin{figure}
\vspace{8.0truecm}
\includegraphics{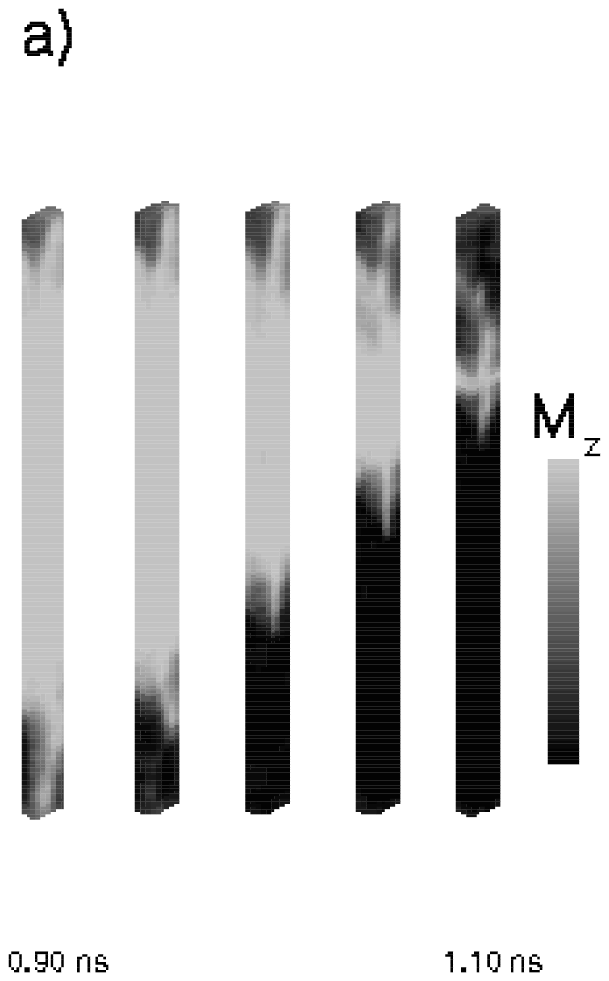}
\includegraphics{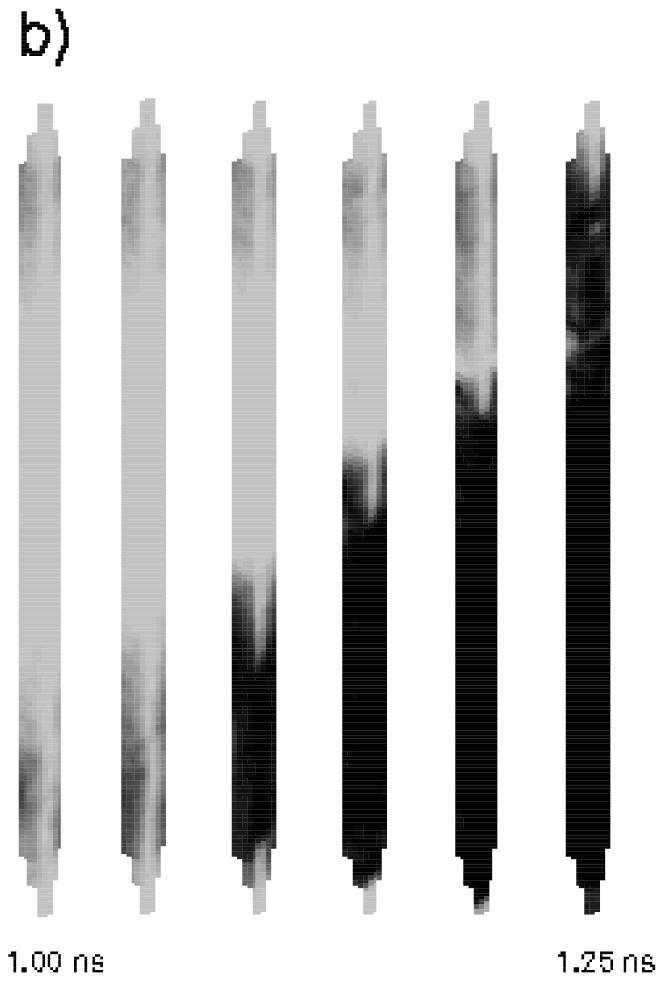}
\caption[]{ Snapshots of the $z$-component of the magnetization,
$M_z$, at equally spaced times for (a): pillars with square ends and
$H_{\rm app}=1800\,{\rm Oe}$ (b): pillars tapered at both ends and
$H_{\rm app}=1975\,{\rm Oe}$.  All are identical cutaway views with
only half of the pillar shown.  Light shades indicate magnetization
pointing in the unfavorable direction while the darkest shades
indicate magnetization pointing in the favorable direction. Both
simulations are for $T=20$~K.  }
\end{figure}

A more dramatic way to see the effects of thermal fluctuations on the
reversal of the magnetization is to hold the magnitude of the applied
field fixed just below the coercive field. This traps the
magnetization in a shallow metastable energy well, and thermal
fluctuations become an essential part of the reversal
process. Snapshots of the $z$-component of the magnetization separated
by $0.05\,{\rm ns}$ are shown in Fig.~2(a). The end caps associated
with pole avoidance become strong in an antiparallel field, and
thermal fluctuations affect their volume. The reversal starts when
thermal fluctuations make one of these regions supercritical
\cite{JAP2000}, after which it propagates at a constant rate towards
the other end of the magnet. Tapering the ends of the magnet reduces
the pole-avoidance effects, and hence affects the end caps. However,
the switching is still characterized by nucleation at the ends; an
example is shown in Fig.~2(b) for pillars $227\,{\rm nm}$ tall. The
major effect we have observed of the tapering is an increase in the
coercive field.

A simple model based on independent nucleation at the ends, followed
by constant growth, allows calculation of the probability of not
switching before time $t$ to be \cite{MMAG,JAP2000}
\begin{equation}
\label{eq:twoexpo}
P_{\rm not}\left(t\right) = 
\left\{
\begin{array}{lr}
  1                                 \qquad\qquad & t< t_0 \\
  e^{-2{\cal I}(t-t_0)} 	    
    \left[1+2{\cal I}(t-t_0)\right] \qquad\qquad & t_0 \le t< 2t_0 \\
  e^{-2{\cal I}(t-t_0)} 	    
    \left[1+2{\cal I}t_0\right]     \qquad\qquad & {2t_0\le t}
\end{array}
\right.
\;,
\end{equation}
where $t_0$ is the earliest time at which a pillar can switch (when
both ends start growing at $t$$=$$0$) and ${\cal I}$ is the nucleation
rate for each end.  This form is compared to the simulation results
for both pillars with flat and tapered ends in Fig.~3. The biggest
obstacle to testing Eq.~(\ref{eq:twoexpo}) is generating sufficient
statistics for the longest and shortest switching
times \cite{MMAG,JAP2000}.

\begin{figure}
\vspace{6.0truecm}
\includegraphics{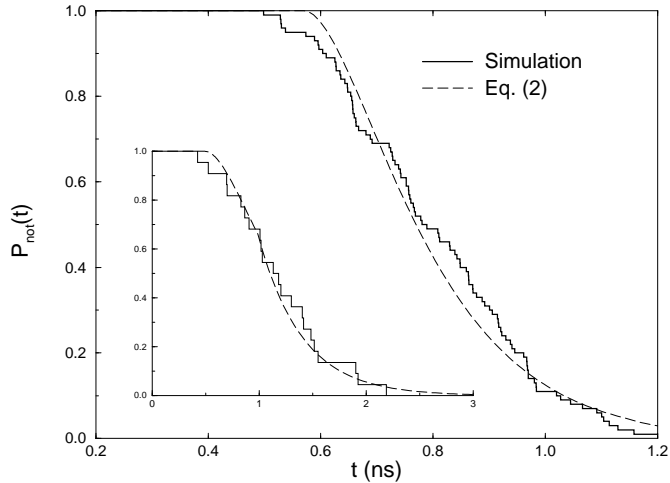}
\caption[] {Probability of not switching, $P_{\rm not}(t)$, for $100$
simulations of the flat-ended pillars at $H_{\rm app}=1800\,{\rm Oe}$
and $T=20\,{\rm K}$, along with a fit to Eq.~(2). The inset shows
similar results for $22$ trials with the tapered-end pillars at
$H_{\rm app}=1975\,{\rm Oe}$ and $T=20\,{\rm K}$.}
\end{figure}

In summary, thermal fluctuations and dipole-dipole interactions have
been included in a three-dimensional model of single-domain nanoscale
magnetic pillars. Simulated dynamics using the Landau-Lifshitz-Gilbert
equation with thermal noise show that magnetization switching in the
pillars occurs as the end caps nucleate, grow, and eventually
coalesce. A simple model based on the thermally-activated nucleation
of end caps for subcoercive fields gives a reasonable description of
the simulation results for $P_{\rm not}(t)$.

Supported by NSF grant No. DMR-9871455, U.S. DOE, NERSC, and by
FSU/SCRI, FSU/CSIT, and FSU/MARTECH.

\end{document}